\documentclass[amsmath,amssymb,twocolumn,prl,showpacs,nofootinbib,floatfix,superscriptaddress]{revtex4-1}
\usepackage{graphics}
\usepackage{dcolumn}
\usepackage {bm}
\usepackage[pdftex]{graphicx}
\usepackage{xcolor}
\usepackage{dcolumn}
\usepackage{footnote}
\usepackage{amssymb}
\usepackage{multirow}
\usepackage{natbib}
\usepackage{xcolor}

\newcommand{\A}[1]{$^{#1}$} 
\newcommand{\pg}{($p,\gamma$)} 
\newcommand{\stater}{\textrm{stat}}
\newcommand{\syser}{\textrm{sys}}

\newcommand{\ket}[1]{\left| #1 \right\rangle}

\begin{document}
\title{Isospin Mixing Reveals \A{30}P\pg\A{31}S Resonance Influencing Nova Nucleosynthesis}
\author{M.~B.~Bennett}
\email{bennettm@nscl.msu.edu}
\affiliation{Department of Physics and Astronomy, Michigan State University, East Lansing, Michigan 48824, USA}
\affiliation{National Superconducting Cyclotron Laboratory, Michigan State University, East Lansing, Michigan 48824, USA}
\affiliation{Joint Institute for Nuclear Astrophysics, Michigan State University, East Lansing, Michigan 48824, USA}
\author{C.~Wrede}
\email{wrede@nscl.msu.edu}
\affiliation{Department of Physics and Astronomy, Michigan State University, East Lansing, Michigan 48824, USA}
\affiliation{National Superconducting Cyclotron Laboratory, Michigan State University, East Lansing, Michigan 48824, USA}
\author{B.~A.~Brown}
\affiliation{Department of Physics and Astronomy, Michigan State University, East Lansing, Michigan 48824, USA}
\affiliation{National Superconducting Cyclotron Laboratory, Michigan State University, East Lansing, Michigan 48824, USA}
\author{S.~N.~Liddick}
\affiliation{National Superconducting Cyclotron Laboratory, Michigan State University, East Lansing, Michigan 48824, USA}
\affiliation{Department of Chemistry, Michigan State University, East Lansing, Michigan 48824, USA}
\author{D.~P\'erez-Loureiro}
\affiliation{Department of Physics and Astronomy, Michigan State University, East Lansing, Michigan 48824, USA}
\affiliation{National Superconducting Cyclotron Laboratory, Michigan State University, East Lansing, Michigan 48824, USA}
\author{D.~W.~Bardayan}
\affiliation{Department of Physics, University of Notre Dame, Notre Dame, Indiana 46556, USA}
\author{A.~A.~Chen}
\affiliation{Department of Physics and Astronomy, McMaster University, Hamilton, Ontario L8S 4M1, Canada}
\author{K.~A.~Chipps}
\affiliation{Oak Ridge National Laboratory, Oak Ridge, Tennessee 37831, USA}
\affiliation{Department of Physics and Astronomy, University of Tennessee, Knoxville, Tennessee 37996, USA}
\author{C.~Fry}
\affiliation{Department of Physics and Astronomy, Michigan State University, East Lansing, Michigan 48824, USA}
\affiliation{National Superconducting Cyclotron Laboratory, Michigan State University, East Lansing, Michigan 48824, USA}
\affiliation{Joint Institute for Nuclear Astrophysics, Michigan State University, East Lansing, Michigan 48824, USA}
\author{B.~E.~Glassman}
\affiliation{Department of Physics and Astronomy, Michigan State University, East Lansing, Michigan 48824, USA}
\affiliation{National Superconducting Cyclotron Laboratory, Michigan State University, East Lansing, Michigan 48824, USA}
\author{C.~Langer}
\affiliation{National Superconducting Cyclotron Laboratory, Michigan State University, East Lansing, Michigan 48824, USA}
\affiliation{Joint Institute for Nuclear Astrophysics, Michigan State University, East Lansing, Michigan 48824, USA}
\author{N.~R.~Larson}
\affiliation{National Superconducting Cyclotron Laboratory, Michigan State University, East Lansing, Michigan 48824, USA}
\affiliation{Department of Chemistry, Michigan State University, East Lansing, Michigan 48824, USA}
\author{E.~I.~McNeice}
\affiliation{Department of Physics and Astronomy, McMaster University, Hamilton, Ontario L8S 4M1, Canada}
\author{Z.~Meisel}
\affiliation{Joint Institute for Nuclear Astrophysics, Michigan State University, East Lansing, Michigan 48824, USA}
\affiliation{Department of Physics, University of Notre Dame, Notre Dame, Indiana 46556, USA}
\author{W.~Ong}
\affiliation{Department of Physics and Astronomy, Michigan State University, East Lansing, Michigan 48824, USA}
\affiliation{National Superconducting Cyclotron Laboratory, Michigan State University, East Lansing, Michigan 48824, USA}
\affiliation{Joint Institute for Nuclear Astrophysics, Michigan State University, East Lansing, Michigan 48824, USA}
\author{P.~D.~O'Malley}
\affiliation{Department of Physics, University of Notre Dame, Notre Dame, Indiana 46556, USA}
\author{S.~D.~Pain}
\affiliation{Oak Ridge National Laboratory, Oak Ridge, Tennessee 37831, USA}
\author{C.~J.~Prokop}
\affiliation{National Superconducting Cyclotron Laboratory, Michigan State University, East Lansing, Michigan 48824, USA}
\affiliation{Department of Chemistry, Michigan State University, East Lansing, Michigan 48824, USA}
\author{H.~Schatz}
\affiliation{Department of Physics and Astronomy, Michigan State University, East Lansing, Michigan 48824, USA}
\affiliation{National Superconducting Cyclotron Laboratory, Michigan State University, East Lansing, Michigan 48824, USA}
\affiliation{Joint Institute for Nuclear Astrophysics, Michigan State University, East Lansing, Michigan 48824, USA}
\author{S.~B.~Schwartz}
\affiliation{Department of Physics and Astronomy, Michigan State University, East Lansing, Michigan 48824, USA}
\affiliation{National Superconducting Cyclotron Laboratory, Michigan State University, East Lansing, Michigan 48824, USA}
\affiliation{Department of Geology and Physics, University of Southern Indiana, Evansville Indiana 47712, USA}
\author{S.~Suchyta}
\affiliation{Department of Chemistry, Michigan State University, East Lansing, Michigan 48824, USA}
\affiliation{National Superconducting Cyclotron Laboratory, Michigan State University, East Lansing, Michigan 48824, USA}
\author{P.~Thompson}
\affiliation{Department of Physics and Astronomy, University of Tennessee, Knoxville, Tennessee 37996, USA}
\author{M.~Walters}
\affiliation{Department of Physics and Astronomy, McMaster University, Hamilton, Ontario L8S 4M1, Canada}
\author{X.~Xu}
\affiliation{Department of Physics and Astronomy, Michigan State University, East Lansing, Michigan 48824, USA}
\affiliation{National Superconducting Cyclotron Laboratory, Michigan State University, East Lansing, Michigan 48824, USA}
%

\begin{abstract}

The thermonuclear \A{30}P\pg\A{31}S reaction rate is critical for modeling the final elemental and isotopic abundances of ONe nova nucleosynthesis, which affect the calibration of proposed nova thermometers and the identification of presolar nova grains, respectively.  Unfortunately, the rate of this reaction is essentially unconstrained experimentally, because the strengths of key \A{31}S proton capture resonance states are not known, largely due to uncertainties in their spins and parities.  Using the $\beta$ decay of \A{31}Cl, we have observed the $\beta$-delayed $\gamma$ decay of a \A{31}S state at $E_x = 6390.2(7)$~keV, with a \A{30}P\pg\A{31}S resonance energy of $E_r = 259.3(8)$~keV, in the middle of the \A{30}P\pg\A{31}S Gamow window for peak nova temperatures.  This state exhibits isospin mixing with the nearby isobaric analog state (IAS) at $E_x = 6279.0(6)$~keV, giving it an unambiguous spin and parity of $3/2^+$ and making it an important $l = 0$ resonance for proton capture on \A{30}P.
\vskip\baselineskip

\noindent \hfill \scriptsize{PACS numbers: 23.20.Lv, 25.40.Lw, 26.30.Ca, 27.30.+t}

\end{abstract}

\maketitle

Inside meteorites retrieved on Earth's surface, grains have been found that exhibit isotopic abundances inconsistent with Solar System abundances.  It is believed that these grains predate the formation of our Solar System; such ``presolar grains'' \cite{ZinnerBOOK2003} likely condensed in the outflows of various stellar sources \cite{JoseApJ2004}.  Through the study of the isotopic compositions of this stardust, a unique branch of astronomy has been developed \cite{ZinnerAREPS1998, DavisPNAS2011}: In-laboratory analysis techniques such as laser ablation and resonant ionization mass spectrometry yield information about the stellar, chemical, and nuclear processes occurring inside extreme astrophysical environments.  However, a grain's stellar origin must first be determined by comparing its measured isotopic ratios with those predicted by astrophysical models.

For example, dust grains are known to condense in the outflows of classical novae \cite{JoseApJ2004}. These thermonuclear explosions, occurring on the surfaces of hydrogen-accreting white-dwarf stars in binary systems \cite{JoseNPA2006}, are crucibles for nucleosynthesis up to $A \sim 40.$  Compared to models for other explosive astrophysical scenarios, the nuclear-physics aspects of nova models are relatively well-understood because most of the essential thermonuclear reaction rates are based on experimental information \cite{IliadisNPA2010}.  However, nucleosynthesis predictions from current hydrodynamic models of oxygen-neon novae are highly uncertain \cite{JoseApJ2004,JoseNPN2005} because the rate of a single reaction, \A{30}P\pg\A{31}S, is essentially unconstrained experimentally.  In fact, the \A{30}P\pg\A{31}S reaction rate commonly employed is derived from the theoretical Hauser-Feshbach statistical model \cite{RauscherADNDT2000}, which is not expected to be accurate at nova temperatures for light nuclides such as \A{31}S, with relatively low densities of states in the region of interest.  The reaction, which is governed by a number of resonances in the region within $\approx$600 keV above the proton emission threshold ($S_p = 6131$ keV \cite{KankainenPRC2010,AME_2012}) at peak nova temperatures of 0.1$-$0.4 GK, is a potential bottleneck in the series of proton captures and $\beta$ decays that characterize nova nucleosynthesis.  \A{30}P\pg\A{31}S competes with the $\beta$ decay of \A{30}P to \A{30}Si ($T_{1/2} = 2.5 \textrm{ min}$), thereby affecting the final abundance ratio \A{30}Si/\A{28}Si and the interpretation of the origins of candidate presolar nova grains based on that ratio \cite{JoseApJ2004}.  If the \A{30}P\pg\A{31}S reaction rate were known, it could also be used to calibrate so-called nova thermometers \cite{DownenAPJ2013}, relationships between model peak temperatures in ONe novae and corresponding simulated elemental abundances that may be compared to abundance observations.

Since sufficiently intense radioactive \A{30}P beams are not yet available for direct measurements of proton captures into the resonant states that govern the reaction rate, indirect methods must be used to populate these important states and measure their properties.  Various experimental probes that have been used in the past include the single-neutron transfer reactions  \A{32}S($p,d$)\A{31}S \cite{MaPRC2007}, \A{32}S($d,t$)\A{31}S \cite{WredePRC2007, IrvinePRC2013}, and \A{32}S$(\textrm{\A{3}He},\alpha)$\A{31}S \cite{VernotteNPA1999}, in-beam $\gamma$-ray spectroscopy measurements of the \A{12}C(\A{20}Ne,$n \gamma$)\A{31}S \cite{JenkinsPRC2005, JenkinsPRC2006}, \A{28}Si(\A{4}He,$n \gamma \gamma$)\A{31}S \cite{DohertyPRL2012,DohertyPRC2014}, \A{24}Mg(\A{16}O,$\alpha \alpha n \gamma $)\A{31}S \cite{VedovaPRC2007,VedovaThesis}, \A{16}O(\A{16}O,$n \gamma$)\A{31}S \cite{PattabiramanPRC2008}, and \A{12}C(\A{20}Ne,$n \gamma$)\A{31}S \cite{TonevJPCS2011} reactions, two measurements of the \A{31}P(\A{3}He,$t$)\A{31}S reaction \cite{WredePRC2007,WredePRC2009,ParikhPRC2011}, and two \A{31}Cl $\beta$-decay experiments \cite{KankainenEPJA2006, SaastamoinenThesis}.

Although it is believed that most of the relevant levels have been populated experimentally \cite{WredeAIP2014, BrownPRC14}, the spin and parity assignments for most of these levels are uncertain and, in many cases, discrepant \cite{ParikhPRC2011, DohertyPRC2014, NDS_A31_2013, BrownPRC14}.  For each of these resonances, spin and parity are needed to determine the resonance strength, which in turn determines the rate of proton capture to that resonance.  Thus, unambiguous spins and parities of resonances in the region of interest are critical for evaluating the \A{30}P\pg\A{31}S reaction rate, predictions of the final abundances of classical novae ejecta, the origin of presolar nova grains, and peak nova temperatures.

The $\beta$ decay of \A{31}Cl preferentially populates $J^\pi = (1/2, 3/2, 5/2)^+ $ states, including the $1/2^+$ and $3/2^+$ states populated by $l = 0$ proton capture on the $J^\pi = 1^+$ \A{30}P nucleus.  These $l = 0$ resonances can have relatively large resonance strengths, since there is no centrifugal barrier impeding proton capture.  The \A{31}Cl $\beta$-decay experiments to date \cite{KankainenEPJA2006, SaastamoinenThesis} have used both $\beta$-delayed proton and $\gamma$ decay through \A{31}S to yield information about astrophysically relevant states.  Despite the relatively low rate of \A{31}Cl production and limited ability for $\gamma$-$\gamma$ coincidence gating in Ref. \cite{SaastamoinenThesis} (and no ability for coincidences in Ref. \cite{KankainenEPJA2006}), both experiments have resulted in the identification of new transitions or levels in \A{31}S; in fact, the isospin $T = 3/2$ isobaric analog state (IAS) of the \A{31}Cl ground state was first definitively identified in Ref. \cite{KankainenEPJA2006} using \A{31}Cl $\beta$-delayed $\gamma$ decay. However, a comparison to shell-model calculations reveals that there are potentially important $J^\pi = (1/2, 3/2, 5/2)^+ $ levels that have not yet been observed in the $\beta$ decay of \A{31}Cl. 

In the present work we report results from a \A{31}Cl $\beta$-delayed $\gamma$-decay experiment using a method similar to Ref. \cite{BennettPRL2013} with significantly improved sensitivity in comparison to Refs. \cite{KankainenEPJA2006, SaastamoinenThesis}.  An intense (maximum 9000 pps), pure (95\%) beam of fast \A{31}Cl ions was produced at the National Superconducting Cyclotron Laboratory (NSCL) at Michigan State University using fragmentation of a 150- MeV/u, 75-pnA \A{36}Ar primary beam from the Coupled Cyclotron Facility incident upon a 1627- mg/cm$^2$ Be transmission target.  Beam purification was accomplished both by magnetic rigidity separation using the A1900 fragment separator \cite{MorrisseyNIM03} and a 145 mg/cm$^2$ Al wedge, and by time-of-flight separation using the Radio Frequency Fragment Separator (RFFS) \cite{BazinNIM09}. Two 300-$\mu$m-thick Si detectors approximately one meter upstream of the experimental setup were lowered periodically into the beam for particle identification purposes.  The main beam contaminants were the radioisotopes \A{24}Na ($\sim$ 2~\% ) and \A{29}P ($\sim$ 1.5\%), with a very small amount of stable \A{28}Si and other lighter ions.  The beam was implanted into a 25-mm-thick plastic scintillator optically coupled to a photomultiplier tube.  Implantations and subsequent $\beta$ decays were detected using the scintillator.  $\beta$-delayed $\gamma$ rays were detected using the Yale Clovershare array: nine high-purity Ge ``clover'' detectors of four crystals each, surrounding the scintillator in two rings of four each, with the ninth detector on the beam axis centered behind the scintillator.  Signals from all 36 clover crystals, the scintillator, and the Si detectors were processed using the NSCL digital data acquisition system \cite{ProkopNIM2014}.

\begin{figure}
\includegraphics[width=\columnwidth]{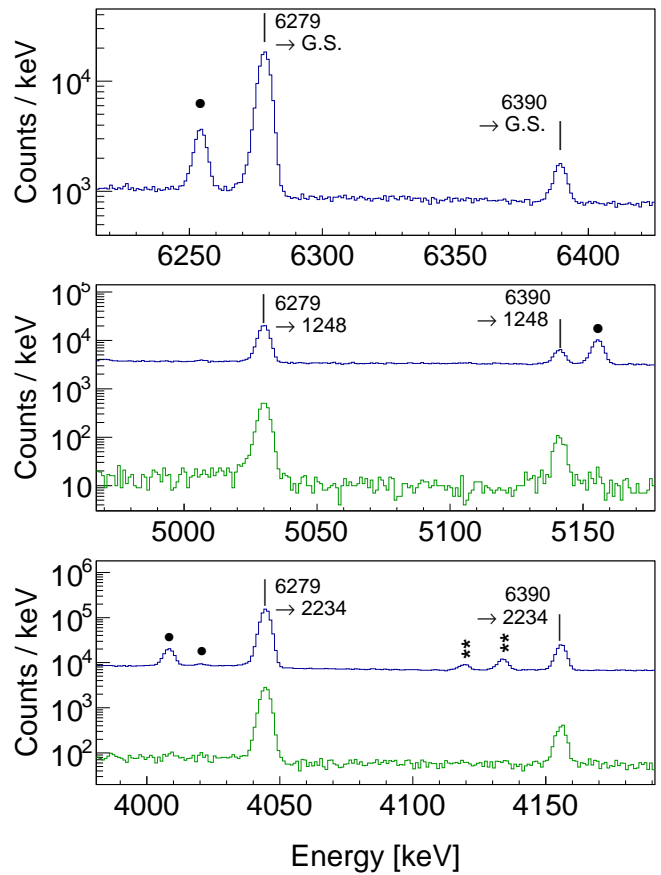} 
\caption{Selected portions of the $\beta$-coincident $\gamma$-ray spectrum [black (blue online) line] showing transitions from the 6279- and 6390-keV \A{31}S states to the ground state and first two excited states ($J^\pi$ = $1/2^+,3/2^+,5/2^+$, respectively).  The bottom two panels also show $\beta$-$\gamma$-$\gamma$ spectra [gray (green online) line] with additional coincidence conditions on the 1248- and 2234-keV $\gamma$ rays, respectively.  Other photopeaks observed from the $\beta$ decay of \A{31}Cl are marked with black circles.  Double escape peaks are marked with double asterisks.}
\label{gamowspectrum}
\end{figure}

In order to facilitate energy and efficiency calibrations, an additional secondary beam of 99\% pure \A{32}Cl was produced using the same primary beam and Al wedge, but with different A1900 and RFFS settings.  Each clover crystal was energy-calibrated using well-known \A{32}Cl $\beta$-delayed $\gamma$-ray peaks up to 7.2 MeV \cite{MelconianPRC12}.  This calibration was applied to a set of \A{31}Cl spectra acquired shortly after the \A{32}Cl spectra, and energy values for the strongest \A{31}Cl peaks were determined.  These peak energies were then used to calibrate and gain-match the entirety of the \A{31}Cl data by treating small portions at a time.  The calibration was checked and verified using an independent cascade-crossover calibration method \cite{KNOLLTEXTBOOK} utilizing only low-energy \A{31}Cl $\beta$-delayed $\gamma$-ray peaks. Systematic uncertainties were approximated by using deviations from literature values of several well-known room background lines and variations in the excitation energies determined using the cascade-crossover method. Systematic uncertainty values were: 0.2 keV for $E_\gamma < 2.7$ MeV, 0.3 keV for 2.7 MeV $ < E_\gamma < 4.8$ MeV, and 0.6 keV for $E_\gamma > 4.8$ MeV.

To perform a relative efficiency calibration for the clover array, the $\gamma$-ray spectrum of a \A{152}Eu calibration source was recorded to produce a relative efficiency curve up to 1400 keV.  A similar curve was also generated using the well-known relative intensities of peaks in the \A{32}Cl data \cite{NDS_A32_2011,MelconianPRC12} from 1547 keV to 7 MeV.  The \A{152}Eu curve was then extrapolated to 1547 keV and the \A{32}Cl curve was scaled to match, producing a continuous relative efficiency curve up to 7 MeV.  Systematic uncertainties in the relative efficiencies included a flat uncertainty of $0.7\% $ at all energies based on variations in the peak-fitting procedure, an uncertainty of 0.2\% for $E_\gamma < 1547$ keV from the \A{152}Eu data, a flat uncertainty of 1.4\% for $E_\gamma$ $>$ 1400 keV  from the uncertainty in the extrapolation of the \A{152}Eu data, and the energy-dependent uncertainty envelope values above 1547 keV in Ref. \cite{MelconianPRC12}: 0.4\% for 1.5 MeV $< E_\gamma <$ 3.5 MeV, 1\% for 3.5 MeV $< E_\gamma <$ 5 MeV, and 5\% for $E_\gamma > 5$ MeV.

To reduce the room background, a cumulative $\beta$-coincident $\gamma$-ray spectrum was produced by requiring coincidences with scintillator events, including $\beta$ decays, in a $1~\mu$s software gate.  Five of the 36 clover crystals were found to have impractically large gain and resolution drifts, so the data from these crystals were discarded.  Thanks to the overall purity of the \A{31}Cl beam, only minimal contributions from beam contaminants were observed.  The ratio of scintillator-gated peak intensity to ungated peak intensity for 18 peaks spanning the $\gamma$-ray energy spectrum was found to have a constant value of 80.6(7)\% for \A{31}Cl, showing that the $\beta$ particle detection efficiency of the scintillator was effectively independent of the $\beta$ end-point energy. The high statistics acquired combined with the high granularity of the Clovershare array also enabled the observation of $\beta$-$\gamma$-$\gamma$ coincidences, which helped to interpret the decay scheme. Samples of the $\beta$-$\gamma$ and $\beta$-$\gamma$-$\gamma$ spectra are shown in Fig. \ref{gamowspectrum}.

\begin{figure}
    \includegraphics[width = \columnwidth]{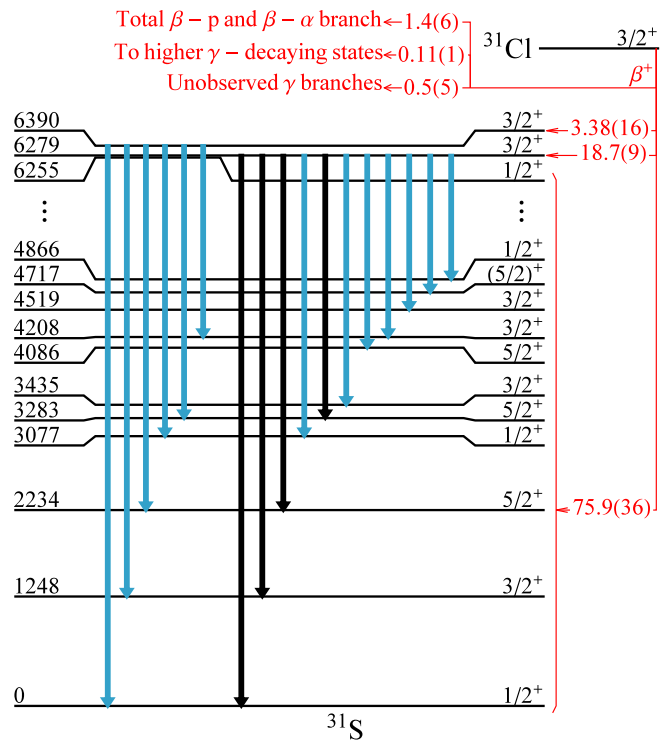}

    \caption{A simplified \A{31}Cl decay scheme focusing on the \A{31}S levels at 6279 (IAS) and 6390 keV. The gray (blue online) vertical arrows indicate previously unobserved transitions.  Energies and intensities for these transitions are listed in Table \ref{energylist}. }
\label{decayschemecomparison}
\end{figure}

$\gamma$-ray energies and intensities were determined by fitting peaks in the $\beta$-coincident $\gamma$-ray spectrum using an exponentially modified Gaussian effective response function.  The peak width and decay constant describing the peak shape were parametrized as a function of energy following Ref. \cite{VarnellNIM69}, ensuring a monotonic variation of peak shape over the 7-MeV range where peaks were fit.  In the $\gamma$-ray spectrum (Fig. \ref{gamowspectrum}), we observed photopeaks corresponding to transitions from three populated \A{31}S states in the region of interest, including the known J$^\pi = 1/2^+$ state at 6255.0(6) keV, the known J$^\pi = 3/2^+, T = 3/2$ IAS at 6279.0(6) keV, and a state at 6390.2(7) keV.  No transitions were observed from states in the excitation energy region between 6390 keV and 7000 keV.  Six transitions from the state at 6390 keV were identified in the $\beta$-$\gamma$ spectrum and all five to excited states were confirmed using $\beta$-$\gamma$-$\gamma$ coincidences (Fig. \ref{gamowspectrum}).  A partial decay scheme is shown in Fig. \ref{decayschemecomparison}. The energies and intensities of these transitions are reported in Table \ref{energylist}.

\begin{table}
\caption{\label{energylist} Energies and intensities of \A{31}Cl$(\beta \gamma)$\A{31}S $\gamma$-ray transitions from the 6279-keV IAS and the state at 6390 keV to other \A{31}S states.  The reported intensities are per 100 $\beta$ decays and the energies have been corrected for nuclear recoil.}

\begin{ruledtabular}
\begin{tabular}{c c c c}
$E_i$ [keV]  & $E_f$ [keV] & $E_{trans}$ [keV] & $I_\gamma$ [\%] \\
 \hline
 $6390.2(7)$                & $0.0$														& $6390.2(6)$											& $0.18(2)$		\\
 $ $												& $1248.4(2)$                     & $5141.7(6)$                     & $0.37(3)$		\\
 $ $    				            & $2234.1(2)$                     & $4156.1(3)$                     & $1.51(7)$	\\
 $ $            				    & $3076.4(3)$                     & $3313.7(3)$                     & $0.40(2)$		\\
 $ $              				  & $3283.8(3)$                     & $3106.4(3)$                     & $0.73(3)$		\\
 $ $             					  & $4207.7(3)$                     & $2182.6(3)$                     & $0.21(1)$		\\
 $ $ & $ $ & $ $ & $ $ \\
 $6279.0(6)$                & $0.0$														& $6278.0(6)$											& $3.2(3)$		\\
 $ $												& $1248.4(2)$                     & $5030.6(6)$                     & $1.9(2)$		\\
 $ $    				            & $2234.1(2)$                     & $4044.9(3)$                     & $11.3(5)$	\\
 $ $            				    & $3076.4(3)$                     & $3202.4(4)$                     & $0.081(6)$		\\
 $ $              				  & $3283.8(3)$                     & $2995.2(3)$                     & $1.15(5)$		\\
 $ $              				  & $3434.9(3)$                     & $2844.0(4)$                     & $0.084(6)$		\\
 $ $              				  & $4085.4(8)$                     & $2192.7(3)$                     & $0.110(8)$		\\
 $ $             					  & $4207.7(3)$                     & $2071.2(2)$                     & $0.58(3)$		\\
 $ $              				  & $4519.6(4)$                     & $1759.1(3)$                     & $0.072(7)$		\\
 $ $              				  & $4717.7(3)$                     & $1561.1(3)$                     & $0.104(7)$		\\
 $ $              				  & $4866.2(6)$                     & $1412.9(3)$                     & $0.082(6)$		\\
\end{tabular}
\end{ruledtabular}
\end{table}

Because the Si detectors used for particle identification were not permanently inserted into the beam, normalizing the $\beta$ feedings of \A{31}S to the total number of implanted \A{31}Cl ions was not the most accurate method available.  Instead, to calculate the $\beta$ feeding for each \A{31}S level populated in the decay of \A{31}Cl, the relative intensity of $\gamma$-ray transitions feeding the level (which for the IAS and the level at 6390 keV was zero) was first subtracted from the relative intensity of $\gamma$-ray transitions deexciting the level.  Then, a 7(2)\% $\beta$ feeding of the \A{31}S ground state following Ref. \cite{KankainenEPJA2006}, a 1.4(6)\% $\beta$-$p$ and $\beta$-$\alpha$  branch based on improvements to the value in Ref. \cite{KankainenEPJA2006} by Ref. \cite{SaastamoinenThesis} and shell-model calculations, and a 0.5(5)\% estimate of unseen $\gamma$ branches based on shell-model calculations were adopted.  Using this sum of 8.9(22)\% for unobserved $\beta$ feeding, the $\beta$ feeding of the observed levels was normalized to the remaining total of 91.1(22)\% and the absolute intensities of the $\gamma$-ray transitions were determined by normalizing to the $\beta$ feedings.  The $\beta$ feeding of the IAS was thus calculated to be $I_{\beta^+} = 18.69 \pm 0.02 (\stater) \pm 0.89 (\syser) \%$, while the $\beta$ feeding of the state at 6390 keV was calculated to be $I_{\beta^+} = 3.38 \pm 0.01 (\stater) \pm 0.15 (\syser) \%$.

The $\beta$ feeding and $\gamma$ branching of the 6390-keV state do not correlate to any state predicted by our shell-model calculations utilizing the universal $sd$-shell version B (USDB) \cite{RichterPRC2008} model.  Furthermore, only the Fermi transition to the IAS would be expected to have such a high $\beta$ feeding at such a high excitation energy, so the possibility of isospin mixing between the state at 6390 keV and the IAS was considered.  Ordinarily, the strength of the Fermi transition $B(F)$ is only nonzero for the transition to the IAS, and in the case of the transition to the IAS from the $T = 3/2, T_z = -3/2$ \A{31}Cl ground state, $B(F) = Z - N = 3$, given that $Z > N$.  However, for the states at 6279 keV and 6390 keV, using the calculated $\beta$ feedings and a $Q$ value based on the \A{31}Cl mass measured in Ref. \cite{KankainenMassPreprint} produces total transition strengths of $B_{6279} = 2.4(1)$ and $B_{6390} = 0.48(3)$.  The inflated transition strength to the level at 6390 keV, the reduced strength to the IAS, and their sum of 2.9(1) are evidence that the Fermi transition is split via isospin mixing, primarily between these two states. By adopting a two-state mixing formalism \cite{TripathiPRL2013}, we deduce an empirical isospin mixing matrix element of 41(1) keV and an unperturbed level spacing of 74(2) keV. Furthermore, we deduce the wave function of the 6390-keV level $\ket{\Psi_{6390}} = 0.913 \ket{T = 1/2} - 0.408 \ket{T = 3/2}$. Recently, strong isospin mixing of a $T = 3/2$ IAS and a $T = 1/2$ state was observed in the $fp$ shell \cite{TripathiPRL2013} and the present work constitutes the first observation of this kind of mixing in the $sd$ shell besides the controversial $A = 23$ case \cite{TripathiPRL2013,IacobA23IAS2006,TigheA23IAS1995}.  The presently observed isospin mixing could also help to explain the recently reported breakdown of the isobaric multiplet mass equation for the $A = 31, T = 3/2$ quartet \cite{KankainenMassPreprint}.

The empirical isospin-mixing values deduced were compared with shell-model calculations that accounted for the mixing of the IAS with all states. With the USDB-cdpn Hamiltonian used in Ref. \cite{BrownPRC14}, as shown in Table I of \cite{BrownPRC14}, there is a triplet of 3/2$^+$ levels with energies of 6205 ($T = 1/2$), 6382 ($T = 1/2$) and 6520 keV ($T = 3/2$) (the energies in Table I of \cite{BrownPRC14} are shifted down by 240 keV compared to those obtained with USDB-cdpn).  The isospin mixing matrix elements are 35 and 12 keV for the first and second of these $T=1/2$ states, respectively. To test the sensitivity of the Hamiltonian, we repeated the USD fit of Ref. \cite{BrownPRC2006} but using only excitation energies (excluding binding energies).  The root-mean-square deviation of 122 keV between theoretical and experimental energies for this fit, called universal $sd$-shell version E (USDE), is similar to that obtained for USDB (126 keV).  The USDE result for $^{31}$S is a triplet of $3/2^+$ states with energies of 6095 ($T = 1/2$), 6184 ($T = 3/2$), and 6375 keV ($T = 1/2$).  The isospin mixing matrix elements are 30 and 27 keV for the first and second $T=1/2$ states, respectively.  Based on the values of the excitation energies and matrix elements for these $3/2^+$ states predicted by USDB and USDE and the theoretical uncertainties implied by their differences, theory is consistent with the present experimental result.  

The experimental results show that the isospin mixing of the IAS is dominated by the 6390-keV state.  The best experimental candidate for the other $T = 1/2, J^\pi = 3/2^+$ level in the triplet predicted by the shell-model calculations is at 5890 keV \cite{DohertyPRL2012,DohertyPRC2014}, and it has an observed $\beta$ feeding of 0.27(2)\%. These values are consistent with the shell-model calculations within theoretical uncertainties.  The relatively small $\beta$ feeding and the relatively large energy difference of this level from the IAS render its isospin mixing with the IAS negligible for the purposes of the present work.

\begin{figure}
    \includegraphics[width = \columnwidth]{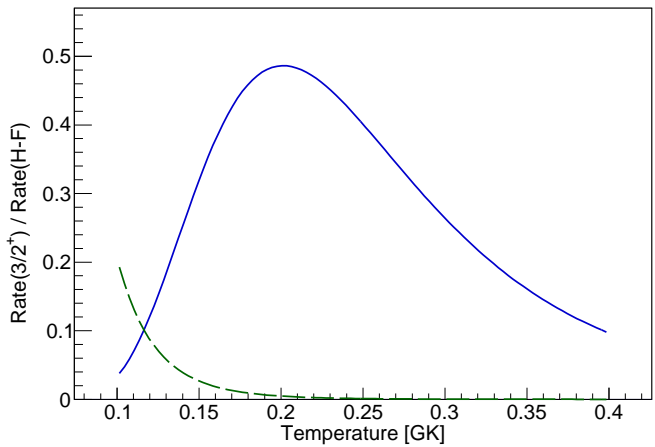}

    \caption{Ratios of the \A{30}P\pg\A{31}S thermonuclear reaction rates calculated for both the new $3/2^+$ state at 6390 keV [solid (blue online) line] and the 6280-keV IAS [dashed (green online) line] to the overall Hauser-Feshbach rate \cite{RauscherADNDT2000}.}
\label{ratecompare}
\end{figure}

The isospin mixing of the IAS and the state at 6390 keV provides a strong, unambiguous constraint on the spin and parity of the 6390 keV state, requiring $J^\pi = 3/2^+$.  This spin and parity make this state an important $l = 0$ resonance for \A{30}P\pg\A{31}S proton capture, located at $E_r = 259.3(8)$ keV, in the heart of the Gamow window for peak nova temperatures.  A spectroscopic factor of 0.0087 was calculated for the unmixed state using USDE and scaled down by the square of the $T = 1/2$ component amplitude ($0.913^2 = 0.83$) to account for the isospin mixing, leading to a proton-decay partial width $\Gamma_p = 36 ~ \mu$eV. This value, combined with the $3/2$ spin of the resonance, gives a \A{30}P\pg\A{31}S resonance strength $\omega \gamma = 24 ~ \mu\textrm{eV}$.

The ratio of the \A{30}P\pg\A{31}S thermonuclear reaction rate calculated at peak nova temperatures using only the 6390-keV resonance to the total Hauser-Feshbach rate \cite{RauscherADNDT2000} is plotted in Fig. \ref{ratecompare}.  Because of the mixing with this $T = 1/2$ state, the 6280-keV IAS also makes a small but non-negligible contribution to the rate, which is also plotted in Fig. \ref{ratecompare}.  The ratio of the 6390-keV state contribution approaches 50\% of the total rate, indicating that this single resonance is very important to the overall \A{30}P\pg\A{31}S resonant capture rate calculation.  It is now the most important \A{30}P\pg\A{31}S resonance with an unambiguous spin and parity identification and, hence, a meaningful estimate of the resonance strength.  Conveniently, the strong population of this resonance in the $\beta$ decay of \A{31}Cl enables measurements of the proton branching ratio, which would yield an experimental value for the resonance strength when combined with measurements of the lifetime.  The relatively large resonance strength may also make this resonance accessible by direct measurements with \A{30}P beams in the future.

The researchers greatly thank the NSCL operations staff for their support and tireless work to ensure the delivery of multiple very pure beams.  This work was supported by the U.S. National Science Foundation under Grants No. PHY-1102511, No. PHY-1404442, No. PHY-1419765, and No. PHY-1431052, the US Department of Energy, National Nuclear Security Administration under Award No. DE-NA0000979, and the Natural Sciences and Engineering Research Council of Canada.

\bibliographystyle{apsrev4-1}
\end{document}